\documentstyle[a4,12pt]{article}
\def\r{{\bf r}}  \def\k{{\bf k}}

\begin{document}



\parindent 0pt
{\large\bf
Quantum Optics in Dispersive and Absorptive Media
}
\bigskip

Akira Shimizu\footnote{
\noindent Corresponding author.
E-mail: shmz@ASone.c.u-tokyo.ac.jp, 
Fax: +81-3-5790-7229
},
Teruaki Okushima and Kazuki Koshino \\
\medskip
Institute of Physics, University of Tokyo,
3-8-1 Komaba, Meguro-ku, Tokyo 153, Japan

\bigskip

{\bf Abstract}
\medskip

Using microscopic models in which both photons and excitons are treated
as microscopic degrees of freedom,
we discuss polaritons of two cases:
One is the case when
excitonic parameters are time dependent.
The time dependence causes creation of
polaritons from a ``false vacuum."
It is shown that both the creation sepctra and the creation efficiency
are much different from the results of the previous studies.
The other is polaritons in absorptive and inhomogeneous cavities.
A polariton in such a system cannot be viewed as a back
and forth oscillation between a photon state and an exciton state.
\bigskip

\bigskip

\parindent 10pt

\noindent{\bf
1.\ Introduction
}\medskip

Optical phenomena in condensed matter are usually discussed
in the case when the optical field can be treated as a {\it classical}
field.
Recently, however,
{\it quantum} optical phenomena in condensed matter,
in which the optical field exhibits its quantum natures in
condensed matter, have been attracting much
attention.
A simple theoretical framework, which is widely used in the literature,
for the analysis of such phenomena
is to somehow quantize  the {\it macroscopic} electromagnetic (EM) fields
in the ``macroscopic Maxwell equations"
(Maxwell equations in matter) \cite{matloob}.
The simple theory treats the matter
as an effective medium, whose properties are assumed to be fully
described by phenomenological parameters such as
a dielectric constant and nonlinear susceptibilities.
The theory works well for simple problems.
However, its validity is not clear
when dispersion and absorption is important,
when the system is inhomogeneous,
when nonlinearities are important,
or when the phenomenological parameters are time dependent.

Another way of formulating quantum optics in condensed matter is
a microscopic approach.
In this case the matter is represented as microscopic
polarization fields, and
both the EM fields and the polarization fields
are quantized.
Upon diagonalization of the Hamiltonian of the
coupled photon-polarization system,
Hopfield \cite{hopfield} obtained
``polaritons" as eigenstates of the total system.
His theory has been extended by many researchers to study polaritons of
various types.
For example, polaritons in the case when excitons
are subject to dissipation were studied by
Huttner, Baumberg and Barnett \cite{huttner}.

In this work, we discuss two cases:
One is polaritons in the case when
excitonic parameters (the exciton energy and exciton-photon
coupling) are time dependent \cite{OS}.
The time dependence causes creation of
polaritons from a ``false vacuum."
This phenomenon is analogous to the dynamical Casimir effect,
which is creation of photons by quick movement of mirrors
in vacuum \cite{fulling,Sa}.
We will present the creation spectra of both
the lower- and upper-branch polaritons.
The other is polaritons in absorptive and inhomogeneous cavities \cite{KS}.
We point out that a polariton in such a system cannot be viewed as a back
and forth oscillation between a photon state and an exciton state, and
a photon state will evolve into a superposition
of many, different exciton states, and will not revive.
\bigskip

\noindent{\bf
2.\ Creation of polaritons from a false vacuum of solids
}\medskip

It is usually assumed in discussions of polaritons that
parameters such as the exciton energy or exciton-photon coupling
are constant, {\it i.e.}, independent of time.
What happens if the parameters are time dependent?
We can show that polaritons will be created even
when the initial state is the vacuum.
This is a general property of quantum theory:
Particles will be created when
the Lagrangian (or, almost equivalently, equations of motion)
has a time-dependent parameter(s) \cite{fulling}.
This can be understood simply as follows.
Suppose that the parameters are constant for $t \leq 0$.
For this constant values of parameters we can (in principle) find out
the ground state, which is called a ``vacuum,"
in which no polaritons are present.
This vacuum is denoted by $| \tilde 0_1 \rangle$.
Assume that the parameters vary during $0 < t < T$,
and again become constant for $t \geq T$.
Then for
$t \geq T$ we can find out another ground state,
which is denoted by $| \tilde 0_2 \rangle$.
Under the initial condition
that the state vector $|\psi \rangle$ is
$|\tilde 0_1 \rangle$ for $t \leq 0$,
we may solve the Schr\"odinger equation \cite{heisenberg}.
Then we will find that
$|\psi \rangle$ at $t \geq T$
differs from $|\tilde 0_2 \rangle$ in general.
This means that we have a finite probability of
finding particles at $t \geq T$.
More particles are normally created for faster variations of
parameters ({\it i.e.}, for shorter $T$).
This is understood if we think of the limiting
case of $T \rightarrow \infty$,
for which  $|\psi \rangle$ evolves adiabatically from
$| \tilde 0_1 \rangle$ at $t \leq 0$
to $|\tilde 0_2 \rangle$ at $t \geq T$,
and no particles are created.
That is, non-adiabatic (fast) variation is necessary for particle
creation.

We here consider the particle creation in condensed matter.
Such a  phenomenon is known in quantum optics as``spontaneous
parametric fluorescence" \cite{yariv}.
The conventional theory of the spontaneous
parametric fluorescence is a phenomenological
one. That is, it relies on the phenomenological quantization
scheme, assuming that dispersion and absorption are absent.
It also assumes a phenomenological interaction Hamiltonian
of the following form;
\begin{equation}
H_{int} = g a_s^\dagger a_i^\dagger a_p + {\rm h.c.},
\end{equation}
where $a_s$, $a_i$ and $a_p$ are annihilation operators of
``signal" (s), ``idler" (i) and pump photons, respectively, and
$g$ is an effective coupling constant which is proportional to
the second-order nonlinear susceptibility.
Through this interaction,
the equations of motion of the signal- and idler-photon fields
are modulated at the frequency $\omega_p$ ($=\omega_s + \omega_i$)
of the pump photon field.
That is, the time scale $T$ of the modulation is
$T \sim 1/\omega_s \sim 1/\omega_i$.
This is a very fast modulation, and
the signal and idler photons can be created efficiently enough to be observed
experimentally.

Similar situations have been studied by Yablonivitch \cite{Ya} and by
Schwinger \cite{Sc}.
They investigated photon creation in the cases when
the dielectric constant $\epsilon$ of material is time dependent.
The time dependence may be due to creation of an electron-hole plasma by
a short laser pulse \cite{Ya},
or by sudden collapse of a bubble in a liquid \cite{Sc}.
As in the case of the spontaneous parametric fluorescence,
the theoretical analyses are phenomenological
ones. That is, they rely on the phenomenological quantization
scheme, assuming that dispersion and absorption are absent.
It was furthermore assumed that $\epsilon$ changes either
discontinuously ($T=0$) \cite{Ya,Sc,sudden},
or almost adiabatically ($T \sim \infty$) \cite{Ya}.
The discontinuous change is unphysical, and, as will be shown below,
turns out to lead to enormous overestimate of the
particle-creation efficiency.

Because of these many assumptions,
the previous theories \cite{yariv,Ya,Sc} have
only a limited range of applicability.
For example, they cannot be applied to
the case when relevant photon energies $\hbar \omega$ are close to
the excitonic energy $\hbar \omega_x$, because the dispersion is strong for
$\omega \sim \omega_x$.
However, such a region of the photon energy
should be most interesting.
Such an interesting region has been studied by
Okushima and Shimizu \cite{OS}, by employing a microscopic model, which is
described by the Lagrangian density of the system of the exciton
field ${\bf X}({\bf r}, t)$, the photon field ${\bf A}({\bf r}, t)$, and
the longitudinal field $U(\r, t)$:
\begin{equation}
{\cal{L}} =
\frac{1}{2} \left[\epsilon_{0}  E^{2}- \frac{1}{\mu_{0} } B^{2} \right]
+
\frac{\rho}{2} \left[ \frac{\partial  X^2}{\partial t}- \omega_x(t)^2 X^2
\right]
-
\alpha (t) \left[ {\bf A}
\cdot \dot{{\bf X}}+U {\bf \nabla} \cdot {\bf X} \right],
\end{equation}
where the electric and magnetic fields are given by
\begin{equation}
{\bf E} = -\dot{\bf A}-{\bf \nabla} U,
\quad
{\bf B} = {\bf \nabla} \times {\bf A}.
\end{equation}
This Lagrangian is a generalization to the time-dependent
$\omega_x$ and $\alpha$ of the model
proposed by Huttner et al.\ \cite{huttner}.
Following them,
we apply the standard quantization procedure as follows.
We impose the Coulomb gauge condition, so that
{\bf A} becomes transversal and $U$ can be eliminated.
By decomposing {\bf X} into the transversal and longitudinal parts,
\begin{equation}
{\bf X} = {\bf X}^\top + {\bf X}^\|,
\end{equation}
we can decompose the Lagrangian $L$
into the transversal and longitudinal parts:
\begin{equation}
L = \int {\cal L} dv
= L^\top + L^\|.
\end{equation}
Since {\bf A} is involved only in $L^\top$, we
focus on this part.
In the Coulomb gauge,
it is convenient to quantize the fields in the $\bf k$ space:
\begin{equation}
{\bf A}(\r,t) = \sum_\k {\bf A} ({\bf k}, t)
e^{i \k \cdot \r},
\quad {\rm where} \quad
{\bf A} ({\bf k}, t) = {\bf A}^\dagger (-{\bf k}, t),
\end{equation}
and similarly for ${\bf X}^\top$.
The fields are further decomposed into two polarizations;
\begin{equation}
{\bf A} ({\bf k}, t) = \sum_{\lambda=1,2} A^\lambda({\bf k},t) {\bf e}_\lambda({\bf k}),
\quad
{\bf k} \cdot {\bf e}_\lambda({\bf k}) = 0,
\end{equation}
and similarly for ${\bf X}^\top$.
By expressing $L^\top$ in terms of these components, we can find
their conjugate momenta $\Pi^\lambda ({\bf k},t)$ and
$P^\lambda ({\bf k},t)$
by differentiating $L^\top$ with respect to
$\dot A^\lambda ({\bf k},t)$ and
$\dot X^\lambda ({\bf k},t)$, respectively.
The Hamiltonian $H$ (for the transversal parts) is obtained as
\begin{equation}
H = \sum_{\k \lambda}
(\Pi^\lambda \dot A^\lambda + P^\lambda \dot X^\lambda) - L^\top.
\end{equation}
Note that this $H$, like $L$, has explicit t dependence,
$H=H(t)$, through
the t dependent parameters.
The fields are then quantized, in the standard manner,
by imposing the equal-time commutation relations;
\begin{equation}
[ A^\lambda({\bf k},t) , \Pi^{\lambda'} ({\bf k}',t) ]
= [ X^\lambda({\bf k},t) , P^{\lambda'} ({\bf k}',t) ]
= i \hbar \delta_{\lambda, \lambda'} \delta_{{\bf k},{\bf k}'}.
\end{equation}
This leads to the Heisenberg equations of motion;
\begin{equation}
i \hbar {\partial \over \partial t} A^\lambda
= [A^\lambda, H(t)],
\quad
i \hbar {\partial \over \partial t} X^\lambda
= [X^\lambda, H(t)].
\label{eqm}\end{equation}

So far, the calculations are parallel to those of Ref.\ \cite{huttner}.
The peculiar features of time-dependent $L$ appear from now on.
Since $H$ is bilinear in the fields, we can diagonalize it
{\it at each $t$\/}.
Let $\tilde{a}^\lambda ({\bf k},t)$ and $\tilde{b}^\lambda ({\bf k},t)$ be
the operators which diagonalize the Hamiltonian {\it at time $t$\/}:
\begin{equation}
  H(t) = \hbar \sum_{{\bf k},\lambda}
\left[
\tilde{\omega}_{a}(k,t) \tilde{a}^{\lambda \dag}({\bf k},t)
\tilde{a}^\lambda ({\bf k},t)+
\tilde{\omega}_{b}(k,t) \tilde{b}^{\lambda \dag}({\bf k},t)
\tilde{b}^\lambda ({\bf k},t)\right] +
\mbox{c-numbers}.
\end{equation}
If $\alpha$ and $\omega_x$ were constant,
$\tilde a^\lambda$ and $\tilde b^\lambda$ would have become the
usual annihilation operators
(times a phase factor, which of course is unimportant)
of lower-branch (LB) and upper-branch (UB) polaritons, respectively,
with $\tilde{\omega}_{a}$ and $\tilde{\omega}_{b}$ being
their eigen-frequencies.
However, this is {\it not \/} the case
because our $\alpha$ and $\omega_x$ depend on $t$.
Moreover, it should also be emphasized that the equations,
\begin{equation}
i \hbar {\partial \over \partial t} \tilde a^\lambda =
[\tilde a^\lambda, H(t)],
\quad
i \hbar {\partial \over \partial t} \tilde b^\lambda =
[\tilde b^\lambda, H(t)],
\quad \mbox{(wrong)}
\end{equation}
which look like Eq.\ (\ref{eqm}),
does {\it not \/} hold for $\tilde a^\lambda$ and $\tilde b^\lambda$.
For these reasons, the diagonalization of $H$
does not solve the problem.

To solve the problem, we must first specify the initial state vector.
For this purpose, we
assume that the parameters are constant in the remote past:
\begin{equation}
\alpha(t) \sim \alpha_1, \ \omega_x(t) \sim \omega_1
\quad {\rm as} \quad t \to -\infty.
\end{equation}
This ensures that as $t \to -\infty$ the operators
$\tilde a^\lambda$ and $\tilde b^\lambda$
and the frequencies $\tilde \omega_a$ and $\tilde \omega_b$
approach the annihilation operators
$\tilde a_1^\lambda$ and $\tilde b_1^\lambda$
and the eigen frequencies
$\tilde \omega_{a1}$ and $\tilde \omega_{b1}$, respectively,
of the usual polariton operators,
which are for the values of the
parameters $\alpha=\alpha_1$ and $\omega_x=\omega_1$.
We consider the case when the initial state vector is the vacuum of
these polaritons:
\begin{equation}
| \psi \rangle = |\tilde 0_1 \rangle,
\quad \mbox{where} \quad
\tilde a_1^\lambda | \tilde 0_1 \rangle =
\tilde b_1^\lambda | \tilde 0_1 \rangle = 0.
\end{equation}
The state vector does not evolve because
we are working in the Heisenberg picture.
On the other hand, to specify the final states (the vacuum,
one-particle states, two-particle states, etc.), we further assume that
the parameters are constant also in the remote future:
\begin{equation}
\alpha(t) \sim \alpha_2, \ \omega_x(t) \sim \omega_2
\quad {\rm as} \quad t \to +\infty.
\end{equation}
Hence, as $t \to +\infty$ the operators
$\tilde a^\lambda$ and $\tilde b^\lambda$
and the frequencies $\tilde \omega_a$ and $\tilde \omega_b$
approach the annihilation operators
$\tilde a_2^\lambda$ and $\tilde b_2^\lambda$
and the eigen frequencies
$\tilde \omega_{a2}$ and $\tilde \omega_{b2}$, respectively,
of another polariton operators,
which are for the values of the
parameters $\alpha=\alpha_2$ and $\omega_x=\omega_2$.

We are interested in the number of
created polaritons of each polariton mode
in the final state.
They are given by
\begin{equation}
\tilde n_{a2}^\lambda (\k) =
\langle \tilde 0_1|
\tilde a_2^{\lambda \dagger}(\k) \tilde a_2^\lambda(\k)
| \tilde 0_1 \rangle,
\quad
\tilde n_{b2}^\lambda (\k) =
\langle \tilde 0_1|
\tilde b_2^{\lambda \dagger}(\k) \tilde b_2^\lambda(\k)
| \tilde 0_1 \rangle,
\end{equation}
for each mode of the LB and UB polaritons, respectively.
To evaluate these numbers, we must express
$\tilde a_2^\lambda$ and $\tilde b_2^\lambda$ in  terms of
$\tilde a_1^\lambda$, $\tilde b_1^\lambda$ and their Hermitian
conjugates.
For our time dependent Lagrangian this relation
takes the form of a Bogoliubov transformation, hence
$\tilde n_{a2}$ and  $\tilde n_{b2}$ become finite \cite{fulling}.
The coefficients of the transformation can be found by
solving the Heisenberg equations of motion for the fields,
Eq.\ (\ref{eqm}),
and inserting the results into the relation between the fields and
$\tilde a^\lambda$, $\tilde b^\lambda$,
$\tilde a^{\lambda \dagger}$ and $\tilde b^{\lambda \dagger}$.

We have performed a numerical calculation
taking the following forms for $\omega_x(t)$ and $\alpha(t)$;
\begin{eqnarray}
\omega_x(t)&=&\left\{
\begin{array}{ll}
\omega_1 & (t \leq 0)\\
\omega_1-
{\omega_1-\omega_2 \over 2 \pi}
[{2 \pi t \over T} - \sin(\frac{2 \pi t}{T})] & (0<t<T) \\
\omega_2 & (t \geq T)\\
\label{singular}\end{array}
\right.\\
\alpha(t)&=&\left\{
\begin{array}{ll}
\alpha_1 & (t \leq 0)\\
\alpha_1-
{\alpha_1-\alpha_2 \over 2 \pi}
[{2 \pi t \over T} - \sin(\frac{2 \pi t}{T})] & (0<t<T) . \\
\alpha_2 & (t \geq T)
\end{array}
\right.
\end{eqnarray}
These are continuous up to the second derivatives,
whereas the third derivatives are discontinuous at
$t = 0$ and $T$.
Figure 1(a) shows the calculated results for the number of created
polaritons of each mode of the lower branch,
$\tilde n_{a2}^\lambda (\k)$, for various values of $\omega_1 T$.
Figure 1(b) shows those of the upper branch,
$\tilde n_{b2}^\lambda (\k)$.
In both cases, strong excitonic features
appear at $c k / \omega_1 \simeq 1$ (which also means
$c k / \omega_2 \simeq 1$ because we have taken $\omega_1 \simeq \omega_2$).
When $c k < \omega_i$ ($c k > \omega_i$), UB (LB)
polaritons are created more
efficiently than LB (UB) polaritons.
This crossover occurs because we are here varying
the exciton parameters,
and the UB polariton has more (less) exciton component
for $c k < \omega_i$ ($c k > \omega_i$).
These characteristics are correctly described only by a microscopic model.

We also find that $\tilde n_{a2}^\lambda$
decreases very quickly as $T$ is increased.
To understand the decrease,
let us investigate the case when
$k = 20 \omega_1/c$ and
$\alpha_1 = \alpha_2 = 0.001 \omega_1 \sqrt{\epsilon_0 \rho}$.
For such large $k$ and small $\alpha$, a LB polariton is
almost an exciton, and we can understand the physics clearly.
The dotted line of Fig.\ 2 plots $\tilde n_{a2}^\lambda$ for this case
as a function of $\omega_1 T$.
We find that  $\tilde n_{a2}^\lambda$ decreases
exponentially as $T$ is increased.
We also find small oscillations for large $T$.
This is found to be due to the fact that our $\omega_x(t)$ has
singularities (third-order derivative is discontinuous
at $t=0$ and $T$).
To verify this we also consider the case when
$\omega_x(t)$ takes the following {\it analytic} form \cite{OS};
\begin{equation}
  \label{analytic}
  \omega^2_x(t) =
  \omega_1^2 +
  {
    (\omega_2^2 - \omega_1^2)
    \over
    1+\exp(-t/\tau)
    },
\end{equation}
where $T=10 \tau$. (The coefficient is chosen
in such a way that a small-T behavior agrees with the
case of the non-analytic $\omega_x$.)
To simplify discussion,
let us take $\alpha=0$, for which
a LB polariton becomes a pure exciton,
and $\tilde n_{a2}^\lambda$ becomes independent of ${\bf k}$.
We then obtain
the exact solution:
\begin{equation}
  \label{exact}
  \tilde n_{a2}^\lambda =
  {
  \sinh^2[\pi (\omega_1 - \omega_2) \tau]
  \over
  \sinh[2 \pi \omega_1 \tau]
  \sinh[2 \pi \omega_2 \tau]
  }.
\end{equation}
This is plotted in Fig.~2 with solid line.
It is seen that the non-analyticity of the form (\ref{singular})
induces oscillations.
It was also shown \cite{OS}
that for large $\omega_1 T$ ($> 20$) the non-analyticity
enhances $\tilde n_{a2}^\lambda$ by
many orders of magnitude.
Since non-analytic change of $\omega_x (t)$ is unphysical,
the result of analytic change of  $\omega_x (t)$ should be more convincing.
However, most previous studies \cite{Ya,Sc,sudden}
assumed that a time-dependent parameter (such as the dielectric constant)
{\it itself\/} has {\it discontinuities\/}.
Our results strongly suggest that
such strong singularities would lead to physically incorrect results.
Regarding our results of Fig.~1,
they may be convincing because our slight singularity
does not play important roles for such short $T$,
as seen from Fig.\ 2.

To summarize this section,
we have considered a polariton system in which
excitonic parameters are time-dependent.
The time dependence causes creation of
polaritons from a ``false vacuum,"
and we have evaluated the number of created polaritons for each polariton
mode.
Since we are interested in the wavelength regions in
which polariton effects are important, and
the dielectric constant exhibits strong dispersions,
we have employed
a microscopic model in which the polarization degrees of freedom are
included as microscopic variables.
Moreover, whereas most previous studies assumed sudden changes
($T=0$) of a parameter(s), we have assumed more realistic situations
in which excitonic parameters vary within a finite time ($T>0$).
Our results strikingly differ from the previous results, both
qualitatively (strong excitonic features at $c k / \omega_1 \simeq 1$) and
quantitatively (by many orders of magnitude).
\bigskip

\noindent{\bf
3.\ Polaritons in absorptive and inhomogeneous cavities
}\medskip

It is usually assumed in discussions of polaritons that
the excitation level,
which couples to photons, of the material is a discrete level.
The excitation level may be
an exciton level in the case of an excitonic polariton,
or a phonon level for a phonon polariton.
As the photon energy is increased to a continuous absorption
spectrum, the single-level approximation breaks down.
Even in such a case, however, we may treat, to a first approximation,
the material excitations as bosons (polarization fields)
because the excitations are composed of bosons or
pair excitations of fermions.
We may therefore write
the Lagrangian density, assuming a one-dimensional system for simplicity,
as \cite{KS}
\begin{eqnarray}
{\cal L} &=& {\cal L}_{EM} + {\cal L}_{mat} +{\cal L}_{int},
\label{total}\\
{\cal L}_{EM} &=& \frac{\epsilon_{0}}{2}
\left[
\dot{A}^{2}-c^{2} \left( \frac{\partial A}{\partial x} \right)^{2}
\right],
\label{em}\\
{\cal L}_{mat} &=& \frac{1}{2} \int_{0}^{\infty} d\omega
\tilde{\rho}_\omega
\left( \dot{\tilde{X}}_{\omega}^{2} - \omega^2 \tilde{X}^{2}_{\omega}
\right),
\label{mat}\\
{\cal L}_{int} &=& - \int_{0}^{\infty} d\omega \tilde{\alpha}_\omega
A\dot{\tilde{X}}_{\omega},
\label{int}
\end{eqnarray}
where $A(x, t)$ is the vector potential of the electromagnetic (EM) field,
and $\tilde{X}_{\omega}(x, t)$ denotes
the polarization field of frequency $\omega$.
Since we are considering the case when
the photon energy lies in a continuous absorption
spectrum, we have taken $\tilde{X}_{\omega}(x, t)$ to have
a continuous label $\omega$, and integral over $\omega$
is performed in Eqs.\
(\ref{mat}) and (\ref{int}).
Moreover, we are interested in the case when the spatial
distribution of the material is inhomogeneous because
in such a case the material constitutes a lossy cavity,
in which behavior of polaritons should be very interesting.
When the material exists only in the regions of $|x| \geq \ell/2$,
we may express this inhomogeneous distribution by imposing
\begin{equation}
\tilde X_{\omega}(x,t)=0 \quad {\rm for} \ |x| < \ell/2.
\end{equation}
The equal-time canonical quantization
of $A$ and $\tilde{X}_{\omega}$ can be performed
in the standard manner, and
we have found that the Hamiltonian is diagonalized as \cite{KS}
\begin{equation}
H =
\sum_{\sigma = \pm}\sum_{q}\int d\omega \hbar\omega
a^{(q\sigma)}(\omega)^{\dag} a^{(q\sigma)}(\omega).
\end{equation}
Here, $a^{(q\sigma)}(\omega)$ is the annihilation operator of a polariton,
which is given by
\begin{eqnarray}
a^{(q\sigma)}(\omega)
&\equiv& \sum_{j}
\left(
 \beta_{j}^{(q\sigma)}(\omega)E_{j\sigma}+
\tilde{\beta}^{(q\sigma)}_{j}(\omega)A_{j\sigma}
\right) \nonumber \\
& + & \sum_{m} \int d\omega^{'}
\left(
 \gamma_{m}^{(q\sigma)}(\omega , \omega^{'})P_{m\sigma , \omega^{'}}
+\tilde{\gamma}^{(q\sigma)}_{m}(\omega , \omega^{'})X_{m\sigma ,\omega^{'}}
\right),
\label{polariton}\end{eqnarray}
where $\sigma=\pm 1$ and $j = 0, 1, 2, \cdots$
(or, $\sigma$ and $q=1, 2, \cdots$)
label some modes of the cavity, $- \epsilon_0 E$ and $P$ are
conjugate momenta of $A$ and $X$, respectively, and $\beta$, $\tilde \beta$,$\gamma$ and $\tilde \gamma$ are some coefficients.
(For details of the notations, see Ref.\ \cite{KS}.)
It is seen that $a^{(q\sigma)}(\omega)$ is a superposition of
infinite number of modes (of $A$ and $X$).
In particular, the integration over continuous exciton modes
results in the breakdown of the standard picture, that
a polariton is a back and forth oscillation of photon
$\to$ exciton $\to$ photon $\to$ exciton $\to \cdots$.
That is, when we prepare a photon state as an initial state,
the state will evolve into a superposition of infinite
number of exciton modes,
and {\it never} returns to the initial photon state.

Using the solution of such unusual polaritons, we can
investigate various physical phenomena in absorptive and inhomogeneous
cavities.
For example, we have evaluated the radiative lifetime of an excited
atom in such a cavity \cite{KS}.
Calculations of other quantities are in progress.

\bigskip\noindent{\bf
Acknowledgment
}\medskip

This work has been supported by the Core Research for Evolutional Science
and Technology (CREST) of the Japan Science and Technology Corporation (JST),
and by Grants-in-Aid for Scientific Research on Priority Areas from the
Ministry of Education, Science and Culture,
and by Sumitomo Foundation.

\bigskip

\begin{center}
{\bf Figure captions}
\end{center}

{\bf Fig.~1.}
        The number of created polaritons per mode
        for (a) lower- and (b) upper-branch polaritons.
        The polariton parameters are taken as
        $\omega_2 = 0.95 \omega_1$,
        $\alpha_1 = 0.05 \omega_1 \sqrt{\epsilon_0 \rho}$,
        $\alpha_2 = 0.03 \omega_1 \sqrt{\epsilon_0 \rho}$,
        and $\omega_1 T = 0.1, 2, 4$.
\bigskip

{\bf Fig.~2.}
        The number of created polaritons per mode for
        lower-branch (LB) polaritons is plotted
        as a function of $\omega_1 T$.
        Dotted line is the result for the LB polariton
        of $k = 20 \omega_1/c$, for
        $\alpha_1 = \alpha_2 = 0.001 \omega_1 \sqrt{\epsilon_0 \rho}$.
        Thin line represents the result when $\omega_x(t)$ takes
        the analytic form, Eq.\ (\ref{analytic}), and
        $\alpha=0$ (which means that a LB polariton in this case
        is a pure exciton).
        In both cases $\omega_2$ is taken to be $0.9 \omega_1$.

\begin{thebibliography}{99}
\bibitem{heisenberg}
Note that in our problem
the Fock space for $t<0$ is different from that for $t>T$
because the parameters which are used to define the Fock
spaces are different.
Hence,
the equivalence of the Scr\"odinger picture and Heisenberg picture
does not necessarily hold (if the system size is infinite).
Although we explain the physics using the Shr\"odinger
picture, the Heisenberg picture is more appropriate
to treat this problem.
Actual calculations have been performed in the Heisenberg
picture.
\bibitem{sudden}
Reference \cite{Sc} introduced a finite time scale $\tau$
which in a sense characterizes time dependence of the dielectric constant.
However, instantaneous changes were assumed at $t=\pm \tau/2$, which
seems to be more relevant in determining the number of created photons.
The same can be said about the model shown in Fig.\ 5 of
Ref.\ \cite{Sa}.
\end{thebibliography}
\end{document}